\documentclass[aps,prl,reprint,superscriptaddress,showpacs]{revtex4-1}
\usepackage{graphicx}
\usepackage{bm}
\usepackage{amsmath}
\usepackage{amssymb}
\usepackage{mathrsfs}
\usepackage{mathtools}
\usepackage{gensymb}
\usepackage{braket}
\usepackage{upgreek}
\usepackage{url}
\usepackage{enumerate}

\usepackage{xcolor}

\usepackage{verbatim}
\usepackage{color}

\usepackage{float}
\usepackage{listings}

\begin{document}

\title{Engineering the Frequency Spectrum of Bright Squeezed Vacuum via Group Velocity Dispersion in an SU(1,1) Interferometer }

\author{Samuel~Lemieux}
\email[]{samzlemieux@gmail.com}
\affiliation{Department of Physics and Max Planck Centre for Extreme and Quantum Photonics, University of Ottawa, 25 Templeton Street, Ottawa, Ontario K1N 6N5, Canada}

\author{Mathieu~Manceau}
\affiliation{Max Planck Institute for the Science of Light, G.-Scharowsky Strasse 1/Bau 24, 91058 Erlangen, Germany}

\author{Polina~R.~Sharapova}
\affiliation{Department of Physics, University of Paderborn, Warburger Strasse 100, D-33098, Paderborn, Germany}
\affiliation{Physics Department, Lomonosov Moscow State University, Moscow 119991, Russia}
\footnotesize
\author{Olga~V.~Tikhonova}
\affiliation{Physics Department, Lomonosov Moscow State University, Moscow 119991, Russia}
\affiliation{Skobeltsyn Institute of Nuclear Physics, Lomonosov Moscow State University, Moscow 119234, Russia}

\author{Robert~W.~Boyd}
\affiliation{Department of Physics and Max Planck Centre for Extreme and Quantum Photonics, University of Ottawa, 25 Templeton Street, Ottawa, Ontario K1N 6N5, Canada}
\affiliation{Institute of Optics, University of Rochester, Rochester, New York 14627, USA}

\author{Gerd~Leuchs}
\affiliation{Max Planck Institute for the Science of Light, G.-Scharowsky Strasse 1/Bau 24, 91058 Erlangen, Germany}
\affiliation{University of Erlangen-Nuremberg, Staudtstrasse 7/B2, 91058 Erlangen, Germany}

\author{Maria~V.~Chekhova}
\affiliation{Max Planck Institute for the Science of Light, G.-Scharowsky Strasse 1/Bau 24, 91058 Erlangen, Germany}
\affiliation{Physics Department, Lomonosov Moscow State University, Moscow 119991, Russia}
\affiliation{University of Erlangen-Nuremberg, Staudtstrasse 7/B2, 91058 Erlangen, Germany}

\date{\today}

\begin{abstract}
Bright squeezed vacuum, a promising tool for quantum information, can be generated by high-gain parametric down-conversion. However, its frequency and angular spectra are typically quite broad, which is undesirable for applications requiring single-mode radiation. We tailor the frequency spectrum of high-gain parametric down-conversion using an SU(1,1) interferometer consisting of two nonlinear crystals with a dispersive medium separating them. The dispersive medium allows us to select a narrow band of the frequency spectrum to be exponentially amplified by high-gain parametric amplification. The frequency spectrum is thereby narrowed from ($56.5 \pm 0.1$) to ($1.22 \pm 0.02$)\,THz and, in doing so, the number of frequency modes is reduced from approximately 50 to $1.82 \pm 0.02$. Moreover, this method provides control and flexibility over the spectrum of the generated light through the timing of the pump.
\end{abstract}

\pacs{03.67.Bg, 03.65.Ud, 42.50.Dv, 42.50.Lc}

\maketitle
\normalsize
Photon pairs have become a ubiquitous tool for many quantum optics applications. They have been used extensively to herald single photons, to increase the phase sensitivity of interferometers and to test the foundations of quantum mechanics through Bell's inequalities~\cite{CAVES1,YURKE1}. A versatile source of biphotons is parametric down-conversion (PDC),  a process in which a pump photon is split in a nonlinear crystal into a pair of entangled photons labeled the signal and the idler. Recent work has featured PDC in the high-gain regime, leading to the generation of bright squeezed vacuum (BSV), that is, squeezed vacuum with a large mean photon number per mode~\cite{chekhova2015bright}. This type of radiation manifests macroscopic quantum features, such as polarization entanglement within a macroscopic pulse~\cite{ISKHAKOV1}, Hong-Ou-Mandel bunching~\cite{ISKHAKOV2}, as well as subshot-noise photon-number correlations~\cite{NRF1,NRF2,NRF3,NRF4,boyer2008entangled,marino2008delocalized,boyer2008generation}. With the detection losses sufficiently small, this state of light will even violate Bell's inequalities~\cite{rosolek2015two}. 

The angular and frequency spectra of PDC are in the general case multimode, with the width inversely proportional to the length of the crystal due to phase matching. At the same time, there are various potential applications that require single-mode BSV, such as phase supersensitivity in SU(1,1) interferometers~\cite{HUDELIST1,YURKE1,ou2012enhancement,plick2010coherent} or heralded preparation of nonclassical states~\cite{harder2016single}. In this Letter, we propose and demonstrate a method to dramatically narrow the frequency spectrum and reduce the number of temporal modes of BSV states. It constitutes a possible extension of the quasisingle transverse mode source of BSV~\cite{PEREZ1}, paving the way for truly single-mode BSV. Unlike most PDC frequency engineering methods proposed to date~\cite{harder2016single,MOSLEY1}, our source makes use of the high gain~(because the desired modes are amplified exponentially), does not require phase-matching engineering, and could be extended to all the transparency range of the nonlinear crystal~\cite{koller2007generation}. Importantly, the method does not involve losses and thus does not destroy photon-number correlations~\cite{PEREZ1}.

The source is based on the interference between the PDC generated by two separated crystals. This scheme, aimed at increasing the parametric gain while at the same time reducing the transverse walk-off~\cite{armstrong1997parametric,chekhova2015bright}, relies on the properties of the SU(1,1) interferometer~\cite{YURKE1}. The latter, in contrast to a usual SU(2) interferometer, has beam splitters replaced by nonlinear media where parametric amplification takes place. In our setup, this leads to the strong selective amplification of some modes and to the deamplification of others.

\begin{figure}
\includegraphics[width=\columnwidth]{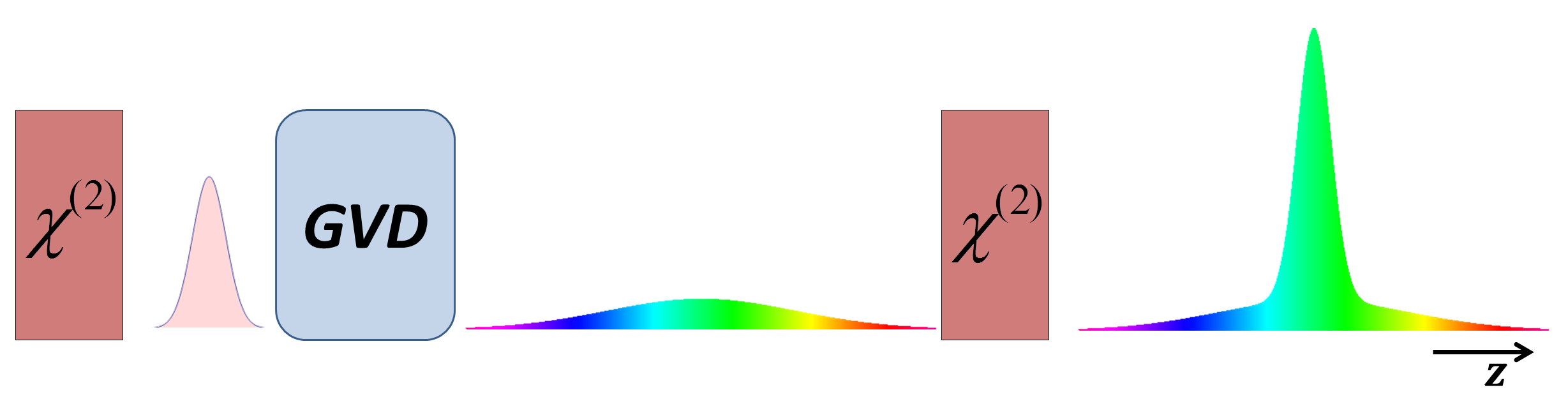}
\caption{Principle of the method. A broadband PDC pulse is generated in a strongly pumped nonlinear crystal; the PDC pulse spreads and chirps after propagation in a medium with group velocity dispersion (GVD); the spread and chirped PDC pulse is amplified in another crystal, using the same pump pulse. The frequency spectrum after the second crystal is narrower than that after the first one. }
\label{fig:GVD}
\end{figure}

By introducing a medium with group velocity dispersion (GVD) between the two crystals forming the SU(1,1) interferometer, we are able to tailor the frequency spectral width of PDC generated in the high-gain regime~(Fig.~\ref{fig:GVD}). A pump pulse sent into the first crystal with quadratic nonlinearity ($\chi ^{(2)}$) generates a broadband PDC pulse. The latter is injected into the GVD medium and undergoes temporal spreading and chirping, meaning that time and frequency contents become correlated. In the second crystal, only the part of the stretched PDC pulse that overlaps in time with the unaltered pump pulse is amplified. This procedure is equivalent to selecting a narrow frequency width without introducing noticeable losses.

The parametric gain $G$ is a key quantity in describing this approach. The average number of output photons per mode $\langle N \rangle$ is given by $\mathrm{sinh}^2(G)$, and $G \propto \chi ^{(2)} E_p L$, where $E_p$ is the pump's electric field amplitude and $L$ is the length of the interaction with the pump pulse, which is given by the crystal length if walk-off effects are negligible.  In the high-gain regime\textemdash large $\chi ^{(2)} E_p L$ product\textemdash doubling the interaction length between the pump pulse and the PDC pulse increases the number of PDC photons by several orders of magnitude. In our case, the interaction length is doubled only for the part of the stretched PDC pulse that overlaps with the unaltered pump pulse in the second crystal, leading to a narrow band output pulse.

Since the PDC pulse is chirped and stretched in time, varying the time of arrival of the pump pulse tunes the central frequency of the amplified PDC. This tuning is performed simply by changing the length of the pump path. In addition, one would expect a twin beam to be created at a frequency determined by the conservation of the pump photon energy. In the spectrum, two peaks would thus appear on either side of the degenerate frequency, and at equal distance therefrom.  

To theoretically describe the interferometer, we assume an undepleted classical pump pulse of fixed duration consisting of plane waves with a Gaussian frequency spectrum of width $\Delta \omega_p$ and central frequency $\bar{\omega}_p$. Moving to the frequency domain and assuming that each photon from the pump pulse spectrum ($p$) gives rise to one signal ($s$) and one idler ($i$) photon, fulfilling $\omega_p=\omega_s+\omega_i$, we obtain a time-independent Hamiltonian. The calculations based on neglecting frequency mismatches are in good agreement with the measurements obtained for the two-crystal interferometer with dispersion~\cite{[A detailed theoretical comparison of the single- and two-crystal configurations in the high-gain regime will be presented in ] Sharapovafreq}, whereas for a single crystal the frequency mismatches have to be taken into account~\cite{quesada2014effects,christ2013theory}.
 
Under these assumptions the Hamiltonian becomes
\begin{equation} \label{eq:hamiltonian}
H = i \hbar \Gamma \int d\omega_s d\omega_i F(\omega_s, \omega_i) a^{\dagger}_{\omega_s} a^{\dagger}_{\omega_i} + \mathrm{H.c.},
\end{equation}
where $\Gamma$ is a measure of the strength of the parametric interaction and $F(\omega_s, \omega_i)$ is called the joint spectral amplitude~(JSA). Following Ref.~\cite{KLYSHKO1}, we find the JSA as
\begin{align} \label{eq:JSA}
F(\omega_s, \omega_i) = \frac{\mathrm{sin}(\delta /2)}{\delta /2} \-\ \mathrm{exp}\bigg[ \frac{- (\omega_s + \omega_i - \overline{\omega}_p)^2}{4 \Delta \omega_p^2} \bigg] \quad \\
\times \-\ \mathrm{exp}[-i (\delta + \delta '/2)] \-\ \mathrm{cos}[(\delta +\delta ')/2] \nonumber ,
\end{align}
for the case of two crystals separated by a linear optical material, where $\delta = [k_p - k_s (\omega_s) - k_i (\omega_i)] L$ is the phase mismatch accumulated in one of the crystals of length $L$ and, likewise, $\delta '$ is the phase mismatch accumulated in the linear gap between the two crystals. The first two factors correspond to the JSA of a single crystal pumped with a pulse of finite duration, assuming conservation of the pump photon's energy. The last two factors convey the total phase acquired by the pump and the PDC in the interferometer, and are responsible for the interference effects appearing in the frequency spectrum. $\delta '$ embodies all the effects of the linear gap between the two crystals, including propagation in different paths, pulse spreading, and chirping.   

To solve the Heisenberg equation of motion for $a^{\dagger}_{\omega_s}$ and $a^{\dagger}_{\omega_i}$, we numerically diagonalize the Hamiltonian by performing a Schmidt decomposition of the JSA~[Eq. (\ref{eq:JSA})] so that it adopts the form
\begin{equation} \label{eq:SVD}
F(\omega_s, \omega_i) = \displaystyle\sum_{k} \sqrt{\lambda_k} u_k(\omega_s) v_k(\omega_i),
\end{equation}
in analogy to Refs.~\cite{SHARAPOVA1,WASILEWSKI,Christ2011}. In the high-gain solution, the weights $\lambda_k$ of the Schmidt (broadband) modes $u_k(\omega_s)$ and $v_k(\omega_i)$ become~\cite{SHARAPOVA1}
\begin{equation}
\tilde{\lambda}_k=\frac{ \mathrm{sinh}^2(G\sqrt{\lambda_k})}{ \displaystyle\sum_{k=1}^{\infty} \mathrm{sinh}^2(G\sqrt{\lambda_k}) },
\label{eq:weights}
\end{equation}
where the parametric gain $G$ is proportional to the coupling strength $\Gamma$. The frequency spectrum of the signal (idler) beam is then found by summing the Schmidt-mode intensity distributions $|u_k(\omega_s)|^2$ ($|v_k(\omega_i)|^2$) with their weights $\tilde{\lambda}_k$. This method provides a complete description of the high-gain PDC spectrum, including the effective number of modes, estimated from the new weights given by Eq.~(\ref{eq:weights}).

In the experiment, the effective number of Schmidt modes $K$ is assessed from the degree of second-order coherence with zero delay $\tau$,
\begin{equation} \label{eq:g2}
g^{(2)}(\tau = 0) = 1 + \frac{g^{(2)}_{1} - 1}{K},
\end{equation}
where $g^{(2)}_{1}$ is the degree of second-order coherence of single-mode PDC~\cite{PhysRevLett.115.143602}. For degenerate collinear PDC, $g^{(2)}_{1} = 3$~\cite{LOUDON1}.  $g^{(2)}$ is obtained experimentally from the definition $g^{(2)}(\tau = 0) = \langle N^2\rangle / \langle N\rangle^2 $, where $N$ is the number of photons in the PDC field~\footnote{By definition, the photon creation and annihilation operators in $g^{(2)}$ should follow normal ordering. When the number of photons is high, it is possible to drop normal ordering.}.

\begin{figure}
\includegraphics[width=0.7 \columnwidth]{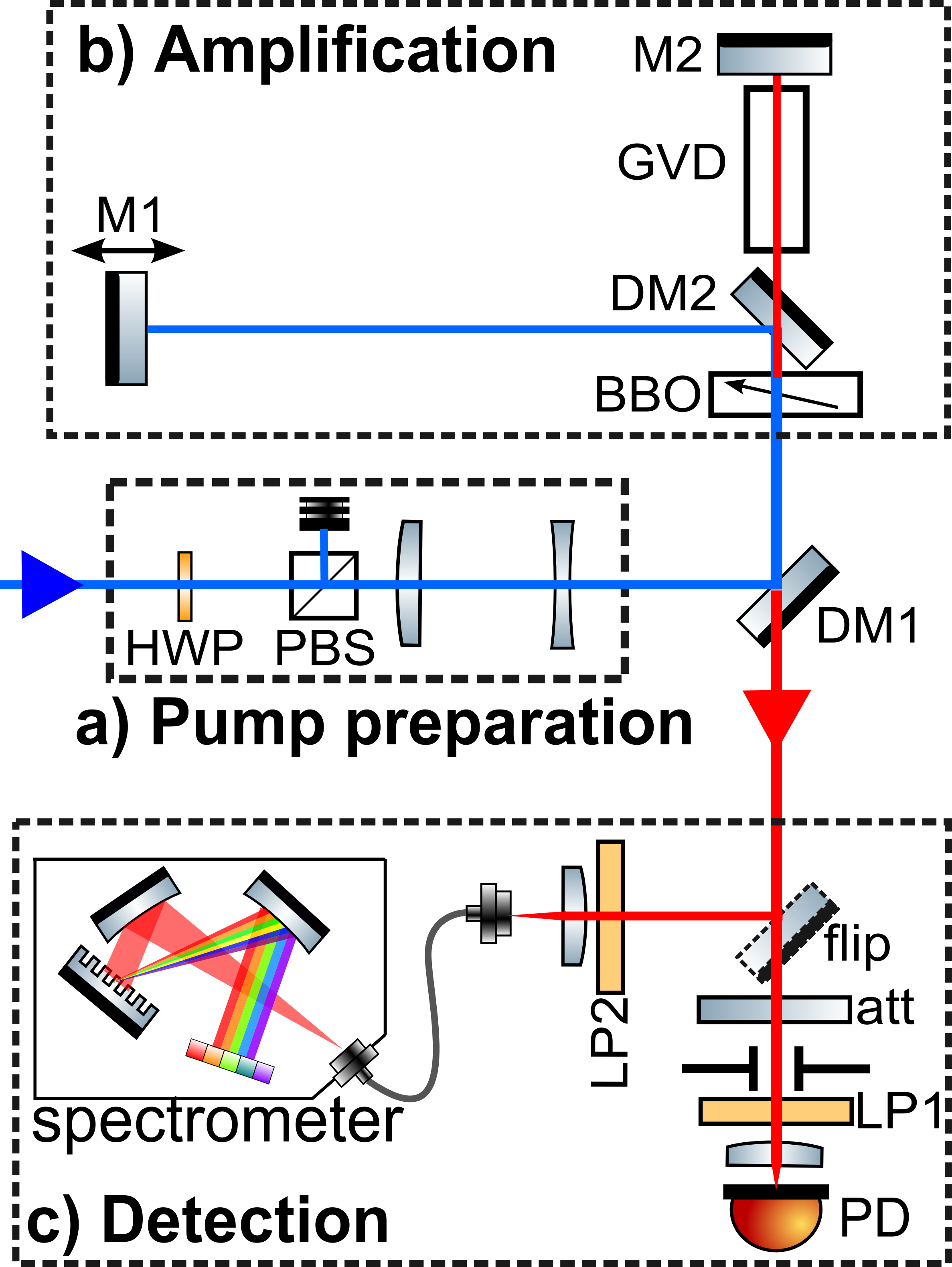}
\caption{Experimental setup. (a) An attenuator [half-wave plate~(HWP) and Glan polarizer~(PBS)] and a telescope are used to prepare the pump in power, polarization and beam diameter. (b) PDC generated in the first pass of the pump in the BBO crystal is stretched temporally in a dispersive medium (GVD) and sent back into the crystal to be amplified. The time of arrival of the pump is carefully tuned with the position of mirror M1, set on a translation stage. (c) The PDC is sent either into a spectrometer or into a photodiode (PD).   }
\label{fig:setup}
\end{figure}
The experimental setup (Fig.~\ref{fig:setup}) consists of an SU(1,1) interferometer of Michelson type~\cite{PhysRevLett.57.2516}. The pump ($\lambda_p$ = 400\,nm) is the second harmonic of a Spectra Physics Spitfire Ace system with a 5\,kHz repetition rate and 0.9\,ps pulses; its power is measured right after an attenuator [half-wave plate (HWP) and a Glan polarizer (PBS)]; and its diameter is set to 0.6\,mm FWHM with a telescope ($f = 300$\,mm and $f = -75$\,mm). A dichroic mirror (DM1, Newlight Photonics HS10-R400/T800) sends the pump towards a nonlinear crystal (BBO, type-I degenerate collinear phase matching, 3\,mm long). Another dichroic mirror (DM2, same as DM1) reflects the pump and transmits the PDC. The pump is reflected by a mirror (M1) set on a translation stage to tune its time of arrival, while the PDC is sent into a dispersive medium (GVD) and reflected back again by a mirror (M2). The PDC and the pump are recombined at DM2. The PDC is amplified in the crystal in accordance with the time of arrival of the pump. After its transmission by DM1, the PDC can either be sent into a low-noise charge integrating detector based on a p-i-n diode (PD) or reflected into the spectrometer by means of a flip mirror (flip). In the detector arm, there is a variable attenuator (att) to prevent saturation, a 500\,$\mu$m pinhole to select a single angular mode, a long-pass filter (LP1) for suppressing the residual pump radiation, and a focusing lens ($f = 40$\,mm). The $p$-$i$-$n$ diode is used to measure the PDC intensity and $g^{(2)}$. The spectrometer arm consists of a long-pass filter (LP2) and a focusing lens ($f = 100$\,mm) to couple light into the spectrometer (Ocean Optics HR4000).  

The dispersive material contains various combinations of SF6, SF57, LLF1, and BK7 glass rods~\cite{glass1}. The amount of pulse spreading is given by the GVD parameter $k'' d$ (dimensionality ps$^2$), where $d$ is the thickness of the material, and  $k'' = d^2 k/d \omega ^2 | _{\omega_0}$ is the GVD evaluated at the central frequency $\omega_0$. For each new rod or combination of rods, one has to set the new position of mirror M1 to provide the proper path length for the pump pulse. The phase shift in the interferometer is not locked and, therefore, drifts randomly, with a typical time of 1 minute. For each set of rods, the pump power is adjusted so that the amplified PDC has the same intensity. This ensures that the parametric gain is kept roughly equal for each measurement. The spectra are recorded only when the interferometer is producing stable output fringes, and the FWHM of the spectra is measured for each value of $k''d$. The theoretical and experimental spectral widths reported herein correspond to the FWHM of spectra in the photon numbers~\footnote{See Supplemental Material at the publisher's website for details on the calibration of the spectrometer and on the effect of phase fluctuations on the shape of the frequency spectrum.}. The number of acquired pulses in $g^{(2)}$ measurements is limited by the drifting time of the interferometer.
\begin{figure}
\includegraphics[width=0.85\columnwidth]{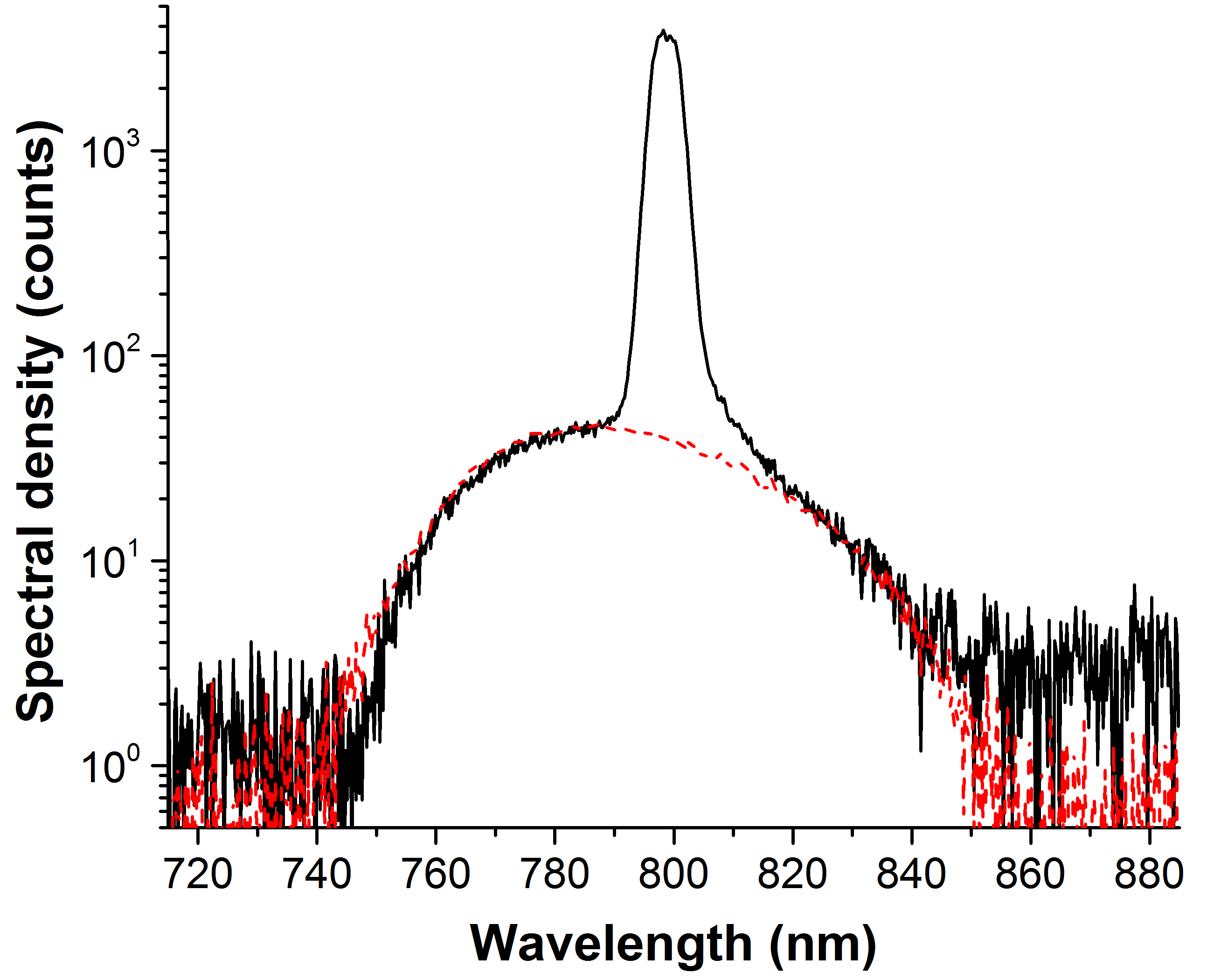}
\caption{Comparison of the spectral densities obtained experimentally for a single pass through the nonlinear crystal (dashed red line), and for the interferometer with a 18.3\,cm rod of SF6 (solid black line).  }
 \label{fig:onecrystal}
\end{figure}

\begin{figure}
\includegraphics[width=0.8\columnwidth]{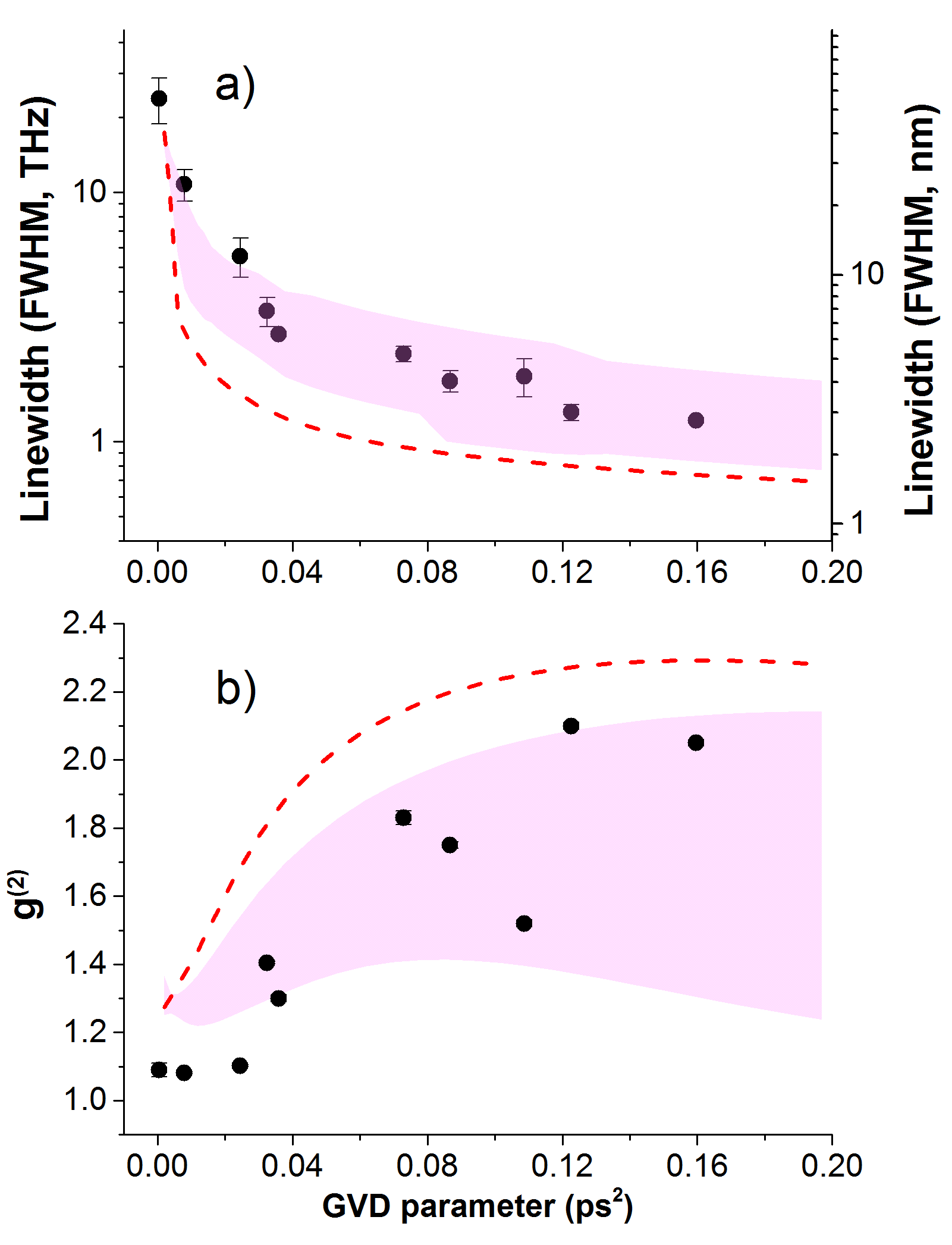}
\caption{Linewidth of the spectrum (a) and $g^{(2)}$ (b) as functions of the GVD parameter. Experimental points (black) are compared to theoretically calculated values (pink zones and red dashed curves). The red dashed curves correspond to constructive interference between the pump and the PDC at degeneracy. The pink zones denote the region within 1 standard deviation from the mean, which is determined by averaging over 8 different phases for the central frequency from 0 to $2\pi$ by varying the air gaps, considering amplification and deamplification cases. The influence of GVD manifests itself through the narrowing of the spectrum and the increase in $g^{(2)}$.}
 \label{fig:widthdisp}
\end{figure}

The introduction of a dispersive material between two nonlinear crystals has the effect of dramatically narrowing the frequency spectrum of PDC, as compared with the spectrum produced by a single crystal (Fig.~\ref{fig:onecrystal}). Moreover, the frequency spectrum of amplified PDC gets narrower as the value of $k''d$ between the two crystals is increased (Fig.~\ref{fig:widthdisp}). The observed asymptotic trend culminates with an experimental spectral width of ($1.22 \pm 0.02$)\,THz, corresponding to the sample with the greatest GVD parameter. The experimental results fit well between the boundaries obtained from theory for different phase scenarios. The effective number of modes $K$, assessed through $g^{(2)}$, is reduced as well when $k''d$ is increased, shrinking down to a minimum of $1.82\pm 0.02$. One would have expected that the points in the right-hand part of Fig.~\ref{fig:widthdisp}, near the asymptote, exhibit single-mode statistics, but the contribution of the phase fluctuations in the interferometer is significant and results in an increase of the number of modes. In fact, phase drifts can transform the single peak of Fig.~\ref{fig:onecrystal} into two or three peaks, effectively changing the frequencies where amplification or deamplification occurs. For smaller GVD values, the spectrum exhibits a rich interference pattern~\cite{Note2, ISKHAKOV3}. The theoretical prediction for a best-case scenario\textemdash phase locked on constructive interference, the dashed curve in Fig.~\ref{fig:widthdisp}\textemdash yields a maximum $g^{(2)}$ of 2.29~($K = 1.55$ modes) with a parametric gain of 7. Under this theory, increasing the gain has the effect of improving the efficiency of the mode selection: the minimum number of modes shrinks to 1.06 when the parametric gain is 10. 

Background multimode PDC, which is not amplified in the second crystal, also has a considerable effect on $g^{(2)}$. As a simple test, we put an interference filter ($\Delta \lambda = 10$\,nm) in front of the detector to cut out part of the background PDC, and  $g^{(2)}$ went from $2.05\pm 0.01$ to $2.29\pm0.02$, corresponding to $K=1.5$. 

\begin{figure}
\includegraphics[width=0.9\columnwidth]{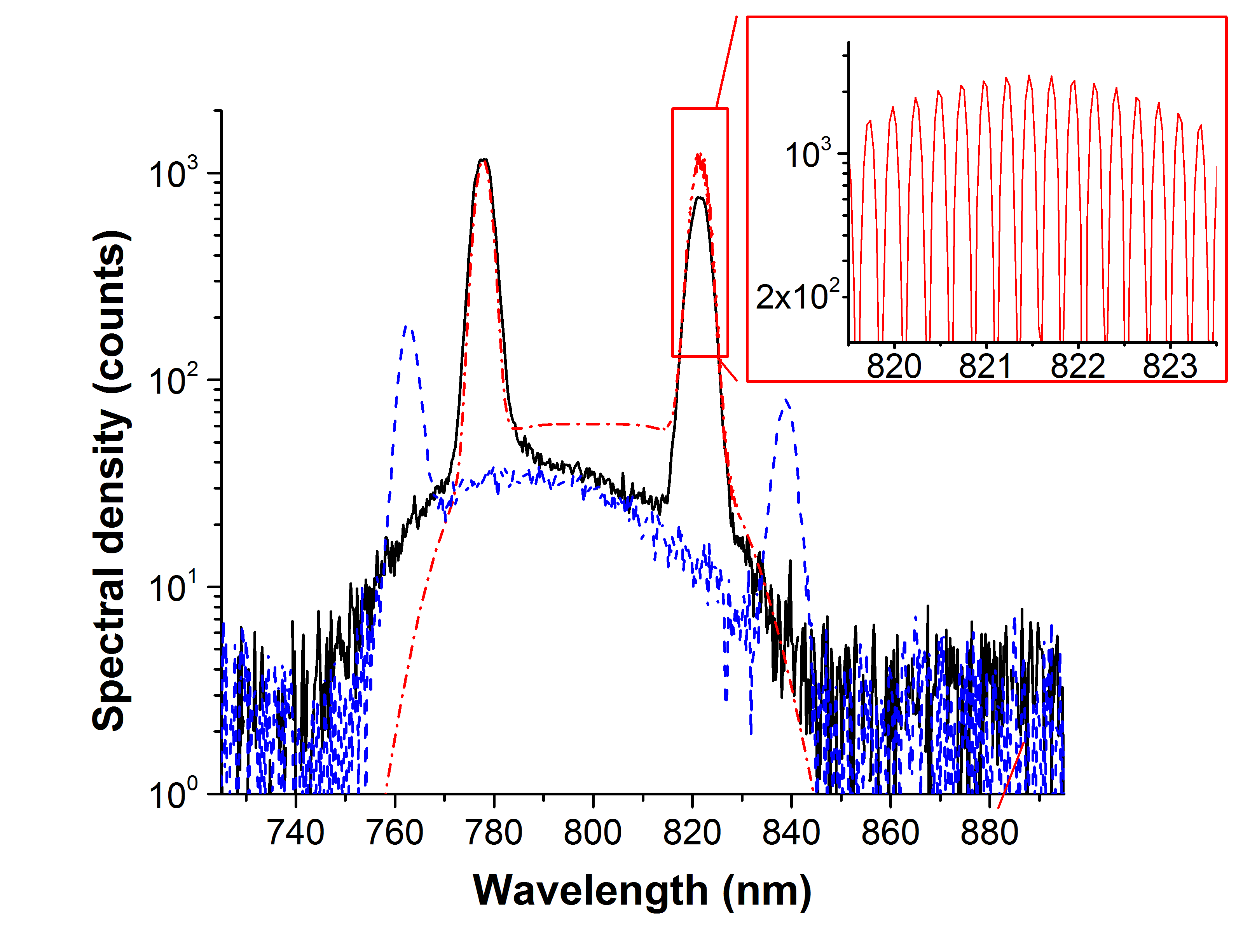}
\caption{Experimental spectra when the pump path length is increased by ($1.40 \pm 0.01$)\,mm~(black solid line) and by ($2.40 \pm 0.01$)\,mm~(blue dashed line) with respect to the configuration for degeneracy. The red dash-dotted line shows the spectrum calculated for the first case, with a parametric gain of 7. The inset shows the second peak of the calculated spectrum without the convolution to show the interference fringes.}
 \label{fig:2color}
\end{figure}

Displacing mirror M1, thereby varying the time of arrival of the pump pulse with respect to the PDC pulse, has the effect of generating two peaks on either side of the degenerate frequency (Fig.~\ref{fig:2color}). A greater mismatch in arrival time induces stronger detuning from degeneracy and weaker PDC intensity, until the two peaks completely fade out. The theory curve of Fig.~\ref{fig:2color} was convolved with a Gaussian whose width corresponds to the nominal resolution of the spectrometer. Without the convolution, one of the two peaks exhibits densely packed fringes~(see the inset in Fig.~\ref{fig:2color}): this interference effect can be understood as induced coherence arising from the indistinguishability of the photon sources~\cite{Sharapovafreq,zou1991induced}. 

In conclusion, by inserting a dispersive medium between two strongly pumped PDC emitters, we have demonstrated BSV with narrowed spectral width, from ($56.5 \pm 0.1$)\,THz down to ($1.22 \pm 0.02$)\,THz. Meanwhile, the number of modes decreased from approximately 50 down to $1.82 \pm 0.02$, and even $1.5$ when the background radiation from a single PDC emitter was eliminated. The method can be used to generate and tune bright narrow band two-color PDC by changing the time of arrival of the pump pulse with respect to the PDC pulse. 

It was earlier demonstrated that by using two spatially separated crystals, one can filter out a single angular mode of PDC~\cite{PEREZ1}. A combination of this method with our approach can be used to generate single-mode BSV in both the frequencies and the angles~\footnote{Note that by selecting a single spatial mode, one does not yet select a single frequency mode, and vice versa.}, but has yet to be demonstrated experimentally. In our case, the pump beam waist was too small to provide the selection of a single spatial mode.

Our method of selective amplification in an SU(1,1) interferometer can be further developed by replacing the GVD material by a pulse shaper imprinting different phases on the modes to be amplified or deamplified. In this manner, one can arbitrarily change the shape of the frequency spectrum amplified in the second crystal. Our source is a significant improvement in the generation of single-mode bright squeezed vacuum states of light. 

\begin{acknowledgments}
The research leading to these results has received funding from the EU FP7 under Grant Agreement No. 308803 (Project BRISQ2). We also acknowledge partial financial support of the Russian Foundation for Basic Research, Grant No. 14-02-00389-a and the joint DFG-RFBR project CH 1591/2-1--16-52-12031 NNIO\_a. This work was also supported by the Canada Excellence Research Chairs program and the National Science and Engineering Research Council of Canada.  We thank Xin Jiang and Patricia Schrehardt for providing the glass samples of SF6, SF57 and LLF1.
\end{acknowledgments}

\bibliography{apstemplate_freqmodes}

\begin{thebibliography}{37}%
\makeatletter
\providecommand \@ifxundefined [1]{%
 \@ifx{#1\undefined}
}%
\providecommand \@ifnum [1]{%
 \ifnum #1\expandafter \@firstoftwo
 \else \expandafter \@secondoftwo
 \fi
}%
\providecommand \@ifx [1]{%
 \ifx #1\expandafter \@firstoftwo
 \else \expandafter \@secondoftwo
 \fi
}%
\providecommand \natexlab [1]{#1}%
\providecommand \enquote  [1]{``#1''}%
\providecommand \bibnamefont  [1]{#1}%
\providecommand \bibfnamefont [1]{#1}%
\providecommand \citenamefont [1]{#1}%
\providecommand \href@noop [0]{\@secondoftwo}%
\providecommand \href [0]{\begingroup \@sanitize@url \@href}%
\providecommand \@href[1]{\@@startlink{#1}\@@href}%
\providecommand \@@href[1]{\endgroup#1\@@endlink}%
\providecommand \@sanitize@url [0]{\catcode `\\12\catcode `\$12\catcode
  `\&12\catcode `\#12\catcode `\^12\catcode `\_12\catcode `\%12\relax}%
\providecommand \@@startlink[1]{}%
\providecommand \@@endlink[0]{}%
\providecommand \url  [0]{\begingroup\@sanitize@url \@url }%
\providecommand \@url [1]{\endgroup\@href {#1}{\urlprefix }}%
\providecommand \urlprefix  [0]{URL }%
\providecommand \Eprint [0]{\href }%
\providecommand \doibase [0]{http://dx.doi.org/}%
\providecommand \selectlanguage [0]{\@gobble}%
\providecommand \bibinfo  [0]{\@secondoftwo}%
\providecommand \bibfield  [0]{\@secondoftwo}%
\providecommand \translation [1]{[#1]}%
\providecommand \BibitemOpen [0]{}%
\providecommand \bibitemStop [0]{}%
\providecommand \bibitemNoStop [0]{.\EOS\space}%
\providecommand \EOS [0]{\spacefactor3000\relax}%
\providecommand \BibitemShut  [1]{\csname bibitem#1\endcsname}%
\let\auto@bib@innerbib\@empty
\bibitem [{\citenamefont {Caves}(1981)}]{CAVES1}%
  \BibitemOpen
  \bibfield  {author} {\bibinfo {author} {\bibfnamefont {C.~M.}\ \bibnamefont
  {Caves}},\ }\href@noop {} {\bibfield  {journal} {\bibinfo  {journal}
  {Physical Review D}\ }\textbf {\bibinfo {volume} {23}},\ \bibinfo {pages}
  {1693} (\bibinfo {year} {1981})}\BibitemShut {NoStop}%
\bibitem [{\citenamefont {Yurke}\ \emph {et~al.}(1986)\citenamefont {Yurke},
  \citenamefont {McCall},\ and\ \citenamefont {Klauder}}]{YURKE1}%
  \BibitemOpen
  \bibfield  {author} {\bibinfo {author} {\bibfnamefont {B.}~\bibnamefont
  {Yurke}}, \bibinfo {author} {\bibfnamefont {S.~L.}\ \bibnamefont {McCall}}, \
  and\ \bibinfo {author} {\bibfnamefont {J.~R.}\ \bibnamefont {Klauder}},\
  }\href@noop {} {\bibfield  {journal} {\bibinfo  {journal} {Physical Review
  A}\ }\textbf {\bibinfo {volume} {33}},\ \bibinfo {pages} {4033} (\bibinfo
  {year} {1986})}\BibitemShut {NoStop}%
\bibitem [{\citenamefont {Chekhova}\ \emph {et~al.}(2015)\citenamefont
  {Chekhova}, \citenamefont {Leuchs},\ and\ \citenamefont
  {{\.Z}ukowski}}]{chekhova2015bright}%
  \BibitemOpen
  \bibfield  {author} {\bibinfo {author} {\bibfnamefont {M.~V.}\ \bibnamefont
  {Chekhova}}, \bibinfo {author} {\bibfnamefont {G.}~\bibnamefont {Leuchs}}, \
  and\ \bibinfo {author} {\bibfnamefont {M.}~\bibnamefont {{\.Z}ukowski}},\
  }\href@noop {} {\bibfield  {journal} {\bibinfo  {journal} {Optics
  Communications}\ }\textbf {\bibinfo {volume} {337}},\ \bibinfo {pages} {27}
  (\bibinfo {year} {2015})}\BibitemShut {NoStop}%
\bibitem [{\citenamefont {Iskhakov}\ \emph {et~al.}(2012)\citenamefont
  {Iskhakov}, \citenamefont {Agafonov}, \citenamefont {Chekhova},\ and\
  \citenamefont {Leuchs}}]{ISKHAKOV1}%
  \BibitemOpen
  \bibfield  {author} {\bibinfo {author} {\bibfnamefont {T.~{\relax Sh}.}\
  \bibnamefont {Iskhakov}}, \bibinfo {author} {\bibfnamefont {I.~N.}\
  \bibnamefont {Agafonov}}, \bibinfo {author} {\bibfnamefont {M.~V.}\
  \bibnamefont {Chekhova}}, \ and\ \bibinfo {author} {\bibfnamefont
  {G.}~\bibnamefont {Leuchs}},\ }\href@noop {} {\bibfield  {journal} {\bibinfo
  {journal} {Physical Review Letters}\ }\textbf {\bibinfo {volume} {109}},\
  \bibinfo {pages} {150502} (\bibinfo {year} {2012})}\BibitemShut {NoStop}%
\bibitem [{\citenamefont {Iskhakov}\ \emph {et~al.}(2013)\citenamefont
  {Iskhakov}, \citenamefont {Spasibko}, \citenamefont {Chekhova},\ and\
  \citenamefont {Leuchs}}]{ISKHAKOV2}%
  \BibitemOpen
  \bibfield  {author} {\bibinfo {author} {\bibfnamefont {T.~{\relax Sh}.}\
  \bibnamefont {Iskhakov}}, \bibinfo {author} {\bibfnamefont {K.~{\relax Yu}.}\
  \bibnamefont {Spasibko}}, \bibinfo {author} {\bibfnamefont {M.~V.}\
  \bibnamefont {Chekhova}}, \ and\ \bibinfo {author} {\bibfnamefont
  {G.}~\bibnamefont {Leuchs}},\ }\href@noop {} {\bibfield  {journal} {\bibinfo
  {journal} {New Journal of Physics}\ }\textbf {\bibinfo {volume} {15}},\
  \bibinfo {pages} {093036} (\bibinfo {year} {2013})}\BibitemShut {NoStop}%
\bibitem [{\citenamefont {Jedrkiewicz}\ \emph {et~al.}(2004)\citenamefont
  {Jedrkiewicz}, \citenamefont {Jiang}, \citenamefont {Brambilla},
  \citenamefont {Gatti}, \citenamefont {Bache}, \citenamefont {Lugiato},\ and\
  \citenamefont {Di~Trapani}}]{NRF1}%
  \BibitemOpen
  \bibfield  {author} {\bibinfo {author} {\bibfnamefont {O.}~\bibnamefont
  {Jedrkiewicz}}, \bibinfo {author} {\bibfnamefont {Y.-K.}\ \bibnamefont
  {Jiang}}, \bibinfo {author} {\bibfnamefont {E.}~\bibnamefont {Brambilla}},
  \bibinfo {author} {\bibfnamefont {A.}~\bibnamefont {Gatti}}, \bibinfo
  {author} {\bibfnamefont {M.}~\bibnamefont {Bache}}, \bibinfo {author}
  {\bibfnamefont {L.~A.}\ \bibnamefont {Lugiato}}, \ and\ \bibinfo {author}
  {\bibfnamefont {P.}~\bibnamefont {Di~Trapani}},\ }\href@noop {} {\bibfield
  {journal} {\bibinfo  {journal} {Physical Review Letters}\ }\textbf {\bibinfo
  {volume} {93}},\ \bibinfo {pages} {243601} (\bibinfo {year}
  {2004})}\BibitemShut {NoStop}%
\bibitem [{\citenamefont {Bondani}\ \emph {et~al.}(2007)\citenamefont
  {Bondani}, \citenamefont {Allevi}, \citenamefont {Zambra}, \citenamefont
  {Paris},\ and\ \citenamefont {Andreoni}}]{NRF2}%
  \BibitemOpen
  \bibfield  {author} {\bibinfo {author} {\bibfnamefont {M.}~\bibnamefont
  {Bondani}}, \bibinfo {author} {\bibfnamefont {A.}~\bibnamefont {Allevi}},
  \bibinfo {author} {\bibfnamefont {G.}~\bibnamefont {Zambra}}, \bibinfo
  {author} {\bibfnamefont {M.~G.~A.}\ \bibnamefont {Paris}}, \ and\ \bibinfo
  {author} {\bibfnamefont {A.}~\bibnamefont {Andreoni}},\ }\href@noop {}
  {\bibfield  {journal} {\bibinfo  {journal} {Physical Review A}\ }\textbf
  {\bibinfo {volume} {76}},\ \bibinfo {pages} {013833} (\bibinfo {year}
  {2007})}\BibitemShut {NoStop}%
\bibitem [{\citenamefont {Brida}\ \emph {et~al.}(2009)\citenamefont {Brida},
  \citenamefont {Caspani}, \citenamefont {Gatti}, \citenamefont {Genovese},
  \citenamefont {Meda},\ and\ \citenamefont {Berchera}}]{NRF3}%
  \BibitemOpen
  \bibfield  {author} {\bibinfo {author} {\bibfnamefont {G.}~\bibnamefont
  {Brida}}, \bibinfo {author} {\bibfnamefont {L.}~\bibnamefont {Caspani}},
  \bibinfo {author} {\bibfnamefont {A.}~\bibnamefont {Gatti}}, \bibinfo
  {author} {\bibfnamefont {M.}~\bibnamefont {Genovese}}, \bibinfo {author}
  {\bibfnamefont {A.}~\bibnamefont {Meda}}, \ and\ \bibinfo {author}
  {\bibfnamefont {I.~R.}\ \bibnamefont {Berchera}},\ }\href@noop {} {\bibfield
  {journal} {\bibinfo  {journal} {Physical Review Letters}\ }\textbf {\bibinfo
  {volume} {102}},\ \bibinfo {pages} {213602} (\bibinfo {year}
  {2009})}\BibitemShut {NoStop}%
\bibitem [{\citenamefont {Agafonov}\ \emph {et~al.}(2010)\citenamefont
  {Agafonov}, \citenamefont {Chekhova},\ and\ \citenamefont {Leuchs}}]{NRF4}%
  \BibitemOpen
  \bibfield  {author} {\bibinfo {author} {\bibfnamefont {I.~N.}\ \bibnamefont
  {Agafonov}}, \bibinfo {author} {\bibfnamefont {M.~V.}\ \bibnamefont
  {Chekhova}}, \ and\ \bibinfo {author} {\bibfnamefont {G.}~\bibnamefont
  {Leuchs}},\ }\href@noop {} {\bibfield  {journal} {\bibinfo  {journal}
  {Physical Review A}\ }\textbf {\bibinfo {volume} {82}},\ \bibinfo {pages}
  {011801} (\bibinfo {year} {2010})}\BibitemShut {NoStop}%
\bibitem [{\citenamefont {Boyer}\ \emph
  {et~al.}(2008{\natexlab{a}})\citenamefont {Boyer}, \citenamefont {Marino},
  \citenamefont {Pooser},\ and\ \citenamefont {Lett}}]{boyer2008entangled}%
  \BibitemOpen
  \bibfield  {author} {\bibinfo {author} {\bibfnamefont {V.}~\bibnamefont
  {Boyer}}, \bibinfo {author} {\bibfnamefont {A.~M.}\ \bibnamefont {Marino}},
  \bibinfo {author} {\bibfnamefont {R.~C.}\ \bibnamefont {Pooser}}, \ and\
  \bibinfo {author} {\bibfnamefont {P.~D.}\ \bibnamefont {Lett}},\ }\href@noop
  {} {\bibfield  {journal} {\bibinfo  {journal} {Science}\ }\textbf {\bibinfo
  {volume} {321}},\ \bibinfo {pages} {544} (\bibinfo {year}
  {2008}{\natexlab{a}})}\BibitemShut {NoStop}%
\bibitem [{\citenamefont {Marino}\ \emph {et~al.}(2008)\citenamefont {Marino},
  \citenamefont {Boyer}, \citenamefont {Pooser}, \citenamefont {Lett},
  \citenamefont {Lemons},\ and\ \citenamefont {Jones}}]{marino2008delocalized}%
  \BibitemOpen
  \bibfield  {author} {\bibinfo {author} {\bibfnamefont {A.~M.}\ \bibnamefont
  {Marino}}, \bibinfo {author} {\bibfnamefont {V.}~\bibnamefont {Boyer}},
  \bibinfo {author} {\bibfnamefont {R.~C.}\ \bibnamefont {Pooser}}, \bibinfo
  {author} {\bibfnamefont {P.~D.}\ \bibnamefont {Lett}}, \bibinfo {author}
  {\bibfnamefont {K.}~\bibnamefont {Lemons}}, \ and\ \bibinfo {author}
  {\bibfnamefont {K.~M.}\ \bibnamefont {Jones}},\ }\href@noop {} {\bibfield
  {journal} {\bibinfo  {journal} {Physical Review Letters}\ }\textbf {\bibinfo
  {volume} {101}},\ \bibinfo {pages} {093602} (\bibinfo {year}
  {2008})}\BibitemShut {NoStop}%
\bibitem [{\citenamefont {Boyer}\ \emph
  {et~al.}(2008{\natexlab{b}})\citenamefont {Boyer}, \citenamefont {Marino},\
  and\ \citenamefont {Lett}}]{boyer2008generation}%
  \BibitemOpen
  \bibfield  {author} {\bibinfo {author} {\bibfnamefont {V.}~\bibnamefont
  {Boyer}}, \bibinfo {author} {\bibfnamefont {A.~M.}\ \bibnamefont {Marino}}, \
  and\ \bibinfo {author} {\bibfnamefont {P.~D.}\ \bibnamefont {Lett}},\
  }\href@noop {} {\bibfield  {journal} {\bibinfo  {journal} {Physical Review
  Letters}\ }\textbf {\bibinfo {volume} {100}},\ \bibinfo {pages} {143601}
  (\bibinfo {year} {2008}{\natexlab{b}})}\BibitemShut {NoStop}%
\bibitem [{\citenamefont {Roso{\l}ek}\ \emph {et~al.}(2015)\citenamefont
  {Roso{\l}ek}, \citenamefont {Stobi{\'n}ska}, \citenamefont {Wie{\'s}niak},\
  and\ \citenamefont {{\.Z}ukowski}}]{rosolek2015two}%
  \BibitemOpen
  \bibfield  {author} {\bibinfo {author} {\bibfnamefont {K.}~\bibnamefont
  {Roso{\l}ek}}, \bibinfo {author} {\bibfnamefont {M.}~\bibnamefont
  {Stobi{\'n}ska}}, \bibinfo {author} {\bibfnamefont {M.}~\bibnamefont
  {Wie{\'s}niak}}, \ and\ \bibinfo {author} {\bibfnamefont {M.}~\bibnamefont
  {{\.Z}ukowski}},\ }\href@noop {} {\bibfield  {journal} {\bibinfo  {journal}
  {Physical Review Letters}\ }\textbf {\bibinfo {volume} {114}},\ \bibinfo
  {pages} {100402} (\bibinfo {year} {2015})}\BibitemShut {NoStop}%
\bibitem [{\citenamefont {Hudelist}\ \emph {et~al.}(2014)\citenamefont
  {Hudelist}, \citenamefont {Kong}, \citenamefont {Liu}, \citenamefont {Jing},
  \citenamefont {Ou},\ and\ \citenamefont {Zhang}}]{HUDELIST1}%
  \BibitemOpen
  \bibfield  {author} {\bibinfo {author} {\bibfnamefont {F.}~\bibnamefont
  {Hudelist}}, \bibinfo {author} {\bibfnamefont {J.}~\bibnamefont {Kong}},
  \bibinfo {author} {\bibfnamefont {C.}~\bibnamefont {Liu}}, \bibinfo {author}
  {\bibfnamefont {J.}~\bibnamefont {Jing}}, \bibinfo {author} {\bibfnamefont
  {Z.}~\bibnamefont {Ou}}, \ and\ \bibinfo {author} {\bibfnamefont
  {W.}~\bibnamefont {Zhang}},\ }\href@noop {} {\bibfield  {journal} {\bibinfo
  {journal} {Nature Communications}\ }\textbf {\bibinfo {volume} {5}},\
  \bibinfo {pages} {3049} (\bibinfo {year} {2014})}\BibitemShut {NoStop}%
\bibitem [{\citenamefont {Ou}(2012)}]{ou2012enhancement}%
  \BibitemOpen
  \bibfield  {author} {\bibinfo {author} {\bibfnamefont {Z.~Y.}\ \bibnamefont
  {Ou}},\ }\href@noop {} {\bibfield  {journal} {\bibinfo  {journal} {Physical
  Review A}\ }\textbf {\bibinfo {volume} {85}},\ \bibinfo {pages} {023815}
  (\bibinfo {year} {2012})}\BibitemShut {NoStop}%
\bibitem [{\citenamefont {Plick}\ \emph {et~al.}(2010)\citenamefont {Plick},
  \citenamefont {Dowling},\ and\ \citenamefont {Agarwal}}]{plick2010coherent}%
  \BibitemOpen
  \bibfield  {author} {\bibinfo {author} {\bibfnamefont {W.~N.}\ \bibnamefont
  {Plick}}, \bibinfo {author} {\bibfnamefont {J.~P.}\ \bibnamefont {Dowling}},
  \ and\ \bibinfo {author} {\bibfnamefont {G.~S.}\ \bibnamefont {Agarwal}},\
  }\href@noop {} {\bibfield  {journal} {\bibinfo  {journal} {New Journal of
  Physics}\ }\textbf {\bibinfo {volume} {12}},\ \bibinfo {pages} {083014}
  (\bibinfo {year} {2010})}\BibitemShut {NoStop}%
\bibitem [{\citenamefont {Harder}\ \emph {et~al.}(2016)\citenamefont {Harder},
  \citenamefont {Bartley}, \citenamefont {Lita}, \citenamefont {Nam},
  \citenamefont {Gerrits},\ and\ \citenamefont
  {Silberhorn}}]{harder2016single}%
  \BibitemOpen
  \bibfield  {author} {\bibinfo {author} {\bibfnamefont {G.}~\bibnamefont
  {Harder}}, \bibinfo {author} {\bibfnamefont {T.~J.}\ \bibnamefont {Bartley}},
  \bibinfo {author} {\bibfnamefont {A.~E.}\ \bibnamefont {Lita}}, \bibinfo
  {author} {\bibfnamefont {S.~W.}\ \bibnamefont {Nam}}, \bibinfo {author}
  {\bibfnamefont {T.}~\bibnamefont {Gerrits}}, \ and\ \bibinfo {author}
  {\bibfnamefont {C.}~\bibnamefont {Silberhorn}},\ }\href@noop {} {\bibfield
  {journal} {\bibinfo  {journal} {Physical Review Letters}\ }\textbf {\bibinfo
  {volume} {116}},\ \bibinfo {pages} {143601} (\bibinfo {year}
  {2016})}\BibitemShut {NoStop}%
\bibitem [{\citenamefont {P{\'e}rez}\ \emph {et~al.}(2014)\citenamefont
  {P{\'e}rez}, \citenamefont {Iskhakov}, \citenamefont {Sharapova},
  \citenamefont {Lemieux}, \citenamefont {Tikhonova}, \citenamefont
  {Chekhova},\ and\ \citenamefont {Leuchs}}]{PEREZ1}%
  \BibitemOpen
  \bibfield  {author} {\bibinfo {author} {\bibfnamefont {A.~M.}\ \bibnamefont
  {P{\'e}rez}}, \bibinfo {author} {\bibfnamefont {T.~{\relax Sh}.}\
  \bibnamefont {Iskhakov}}, \bibinfo {author} {\bibfnamefont {P.}~\bibnamefont
  {Sharapova}}, \bibinfo {author} {\bibfnamefont {S.}~\bibnamefont {Lemieux}},
  \bibinfo {author} {\bibfnamefont {O.~V.}\ \bibnamefont {Tikhonova}}, \bibinfo
  {author} {\bibfnamefont {M.~V.}\ \bibnamefont {Chekhova}}, \ and\ \bibinfo
  {author} {\bibfnamefont {G.}~\bibnamefont {Leuchs}},\ }\href@noop {}
  {\bibfield  {journal} {\bibinfo  {journal} {Optics Letters}\ }\textbf
  {\bibinfo {volume} {39}},\ \bibinfo {pages} {2403} (\bibinfo {year}
  {2014})}\BibitemShut {NoStop}%
\bibitem [{\citenamefont {Mosley}\ \emph {et~al.}(2008)\citenamefont {Mosley},
  \citenamefont {Lundeen}, \citenamefont {Smith}, \citenamefont {Wasylczyk},
  \citenamefont {U'Ren}, \citenamefont {Silberhorn},\ and\ \citenamefont
  {Walmsley}}]{MOSLEY1}%
  \BibitemOpen
  \bibfield  {author} {\bibinfo {author} {\bibfnamefont {P.~J.}\ \bibnamefont
  {Mosley}}, \bibinfo {author} {\bibfnamefont {J.~S.}\ \bibnamefont {Lundeen}},
  \bibinfo {author} {\bibfnamefont {B.~J.}\ \bibnamefont {Smith}}, \bibinfo
  {author} {\bibfnamefont {P.}~\bibnamefont {Wasylczyk}}, \bibinfo {author}
  {\bibfnamefont {A.~B.}\ \bibnamefont {U'Ren}}, \bibinfo {author}
  {\bibfnamefont {C.}~\bibnamefont {Silberhorn}}, \ and\ \bibinfo {author}
  {\bibfnamefont {I.~A.}\ \bibnamefont {Walmsley}},\ }\href@noop {} {\bibfield
  {journal} {\bibinfo  {journal} {Physical Review Letters}\ }\textbf {\bibinfo
  {volume} {100}},\ \bibinfo {pages} {133601} (\bibinfo {year}
  {2008})}\BibitemShut {NoStop}%
\bibitem [{\citenamefont {Koller}\ \emph {et~al.}(2007)\citenamefont {Koller},
  \citenamefont {Haiser}, \citenamefont {Huber}, \citenamefont {Schrader},
  \citenamefont {Regner}, \citenamefont {Schreier},\ and\ \citenamefont
  {Zinth}}]{koller2007generation}%
  \BibitemOpen
  \bibfield  {author} {\bibinfo {author} {\bibfnamefont {F.~O.}\ \bibnamefont
  {Koller}}, \bibinfo {author} {\bibfnamefont {K.}~\bibnamefont {Haiser}},
  \bibinfo {author} {\bibfnamefont {M.}~\bibnamefont {Huber}}, \bibinfo
  {author} {\bibfnamefont {T.~E.}\ \bibnamefont {Schrader}}, \bibinfo {author}
  {\bibfnamefont {N.}~\bibnamefont {Regner}}, \bibinfo {author} {\bibfnamefont
  {W.~J.}\ \bibnamefont {Schreier}}, \ and\ \bibinfo {author} {\bibfnamefont
  {W.}~\bibnamefont {Zinth}},\ }\href@noop {} {\bibfield  {journal} {\bibinfo
  {journal} {Optics Letters}\ }\textbf {\bibinfo {volume} {32}},\ \bibinfo
  {pages} {3339} (\bibinfo {year} {2007})}\BibitemShut {NoStop}%
\bibitem [{\citenamefont {Armstrong}\ \emph {et~al.}(1997)\citenamefont
  {Armstrong}, \citenamefont {Alford}, \citenamefont {Raymond}, \citenamefont
  {Smith},\ and\ \citenamefont {Bowers}}]{armstrong1997parametric}%
  \BibitemOpen
  \bibfield  {author} {\bibinfo {author} {\bibfnamefont {D.~J.}\ \bibnamefont
  {Armstrong}}, \bibinfo {author} {\bibfnamefont {W.~J.}\ \bibnamefont
  {Alford}}, \bibinfo {author} {\bibfnamefont {T.~D.}\ \bibnamefont {Raymond}},
  \bibinfo {author} {\bibfnamefont {A.~V.}\ \bibnamefont {Smith}}, \ and\
  \bibinfo {author} {\bibfnamefont {M.~S.}\ \bibnamefont {Bowers}},\
  }\href@noop {} {\bibfield  {journal} {\bibinfo  {journal} {JOSA B}\ }\textbf
  {\bibinfo {volume} {14}},\ \bibinfo {pages} {460} (\bibinfo {year}
  {1997})}\BibitemShut {NoStop}%
\bibitem [{\citenamefont {Sharapova}\ \emph {et~al.}(tion)\citenamefont
  {Sharapova}, \citenamefont {Tikhonova}, \citenamefont {Lemieux},
  \citenamefont {Boyd}, \citenamefont {Leuchs},\ and\ \citenamefont
  {Chekhova}}]{Sharapovafreq}%
  \BibitemOpen
  \bibfield  {author} {\bibinfo {author} {\bibfnamefont {P.}~\bibnamefont
  {Sharapova}}, \bibinfo {author} {\bibfnamefont {O.~V.}\ \bibnamefont
  {Tikhonova}}, \bibinfo {author} {\bibfnamefont {S.}~\bibnamefont {Lemieux}},
  \bibinfo {author} {\bibfnamefont {R.~W.}\ \bibnamefont {Boyd}}, \bibinfo
  {author} {\bibfnamefont {G.}~\bibnamefont {Leuchs}}, \ and\ \bibinfo {author}
  {\bibfnamefont {M.~V.}\ \bibnamefont {Chekhova}},\ }\href@noop {} {\
  (\bibinfo {year} {in preparation})}\BibitemShut {NoStop}%
\bibitem [{\citenamefont {Quesada}\ and\ \citenamefont
  {Sipe}(2014)}]{quesada2014effects}%
  \BibitemOpen
  \bibfield  {author} {\bibinfo {author} {\bibfnamefont {N.}~\bibnamefont
  {Quesada}}\ and\ \bibinfo {author} {\bibfnamefont {J.~E.}\ \bibnamefont
  {Sipe}},\ }\href@noop {} {\bibfield  {journal} {\bibinfo  {journal} {Physical
  Review A}\ }\textbf {\bibinfo {volume} {90}},\ \bibinfo {pages} {063840}
  (\bibinfo {year} {2014})}\BibitemShut {NoStop}%
\bibitem [{\citenamefont {Christ}\ \emph {et~al.}(2013)\citenamefont {Christ},
  \citenamefont {Brecht}, \citenamefont {Mauerer},\ and\ \citenamefont
  {Silberhorn}}]{christ2013theory}%
  \BibitemOpen
  \bibfield  {author} {\bibinfo {author} {\bibfnamefont {A.}~\bibnamefont
  {Christ}}, \bibinfo {author} {\bibfnamefont {B.}~\bibnamefont {Brecht}},
  \bibinfo {author} {\bibfnamefont {W.}~\bibnamefont {Mauerer}}, \ and\
  \bibinfo {author} {\bibfnamefont {C.}~\bibnamefont {Silberhorn}},\
  }\href@noop {} {\bibfield  {journal} {\bibinfo  {journal} {New Journal of
  Physics}\ }\textbf {\bibinfo {volume} {15}},\ \bibinfo {pages} {053038}
  (\bibinfo {year} {2013})}\BibitemShut {NoStop}%
\bibitem [{\citenamefont {Klyshko}(1993)}]{KLYSHKO1}%
  \BibitemOpen
  \bibfield  {author} {\bibinfo {author} {\bibfnamefont {D.}~\bibnamefont
  {Klyshko}},\ }\href@noop {} {\bibfield  {journal} {\bibinfo  {journal}
  {Soviet Journal of Experimental and Theoretical Physics}\ }\textbf {\bibinfo
  {volume} {77}},\ \bibinfo {pages} {222} (\bibinfo {year} {1993})}\BibitemShut
  {NoStop}%
\bibitem [{\citenamefont {Sharapova}\ \emph {et~al.}(2015)\citenamefont
  {Sharapova}, \citenamefont {P\'e­rez}, \citenamefont {Tikhonova},\ and\
  \citenamefont {Chekhova}}]{SHARAPOVA1}%
  \BibitemOpen
  \bibfield  {author} {\bibinfo {author} {\bibfnamefont {P.}~\bibnamefont
  {Sharapova}}, \bibinfo {author} {\bibfnamefont {A.~M.}\ \bibnamefont
  {P\'e­rez}}, \bibinfo {author} {\bibfnamefont {O.~V.}\ \bibnamefont
  {Tikhonova}}, \ and\ \bibinfo {author} {\bibfnamefont {M.~V.}\ \bibnamefont
  {Chekhova}},\ }\href@noop {} {\bibfield  {journal} {\bibinfo  {journal}
  {Physical Review A}\ }\textbf {\bibinfo {volume} {91}},\ \bibinfo {pages}
  {043816} (\bibinfo {year} {2015})}\BibitemShut {NoStop}%
\bibitem [{\citenamefont {Wasilewski}\ \emph {et~al.}(2006)\citenamefont
  {Wasilewski}, \citenamefont {Lvovsky}, \citenamefont {Banaszek},\ and\
  \citenamefont {Radzewicz}}]{WASILEWSKI}%
  \BibitemOpen
  \bibfield  {author} {\bibinfo {author} {\bibfnamefont {W.}~\bibnamefont
  {Wasilewski}}, \bibinfo {author} {\bibfnamefont {A.~I.}\ \bibnamefont
  {Lvovsky}}, \bibinfo {author} {\bibfnamefont {K.}~\bibnamefont {Banaszek}}, \
  and\ \bibinfo {author} {\bibfnamefont {C.}~\bibnamefont {Radzewicz}},\
  }\href@noop {} {\bibfield  {journal} {\bibinfo  {journal} {Physical Review
  A}\ }\textbf {\bibinfo {volume} {73}},\ \bibinfo {pages} {063819} (\bibinfo
  {year} {2006})}\BibitemShut {NoStop}%
\bibitem [{\citenamefont {Christ}\ \emph {et~al.}(2011)\citenamefont {Christ},
  \citenamefont {Laiho}, \citenamefont {Eckstein}, \citenamefont {Cassemiro},\
  and\ \citenamefont {Silberhorn}}]{Christ2011}%
  \BibitemOpen
  \bibfield  {author} {\bibinfo {author} {\bibfnamefont {A.}~\bibnamefont
  {Christ}}, \bibinfo {author} {\bibfnamefont {K.}~\bibnamefont {Laiho}},
  \bibinfo {author} {\bibfnamefont {A.}~\bibnamefont {Eckstein}}, \bibinfo
  {author} {\bibfnamefont {K.~N.}\ \bibnamefont {Cassemiro}}, \ and\ \bibinfo
  {author} {\bibfnamefont {C.}~\bibnamefont {Silberhorn}},\ }\href
  {http://stacks.iop.org/1367-2630/13/i=3/a=033027} {\bibfield  {journal}
  {\bibinfo  {journal} {New Journal of Physics}\ }\textbf {\bibinfo {volume}
  {13}},\ \bibinfo {pages} {033027} (\bibinfo {year} {2011})}\BibitemShut
  {NoStop}%
\bibitem [{\citenamefont {Finger}\ \emph {et~al.}(2015)\citenamefont {Finger},
  \citenamefont {Iskhakov}, \citenamefont {Joly}, \citenamefont {Chekhova},\
  and\ \citenamefont {Russell}}]{PhysRevLett.115.143602}%
  \BibitemOpen
  \bibfield  {author} {\bibinfo {author} {\bibfnamefont {M.~A.}\ \bibnamefont
  {Finger}}, \bibinfo {author} {\bibfnamefont {T.~{\relax Sh}.}\ \bibnamefont
  {Iskhakov}}, \bibinfo {author} {\bibfnamefont {N.~Y.}\ \bibnamefont {Joly}},
  \bibinfo {author} {\bibfnamefont {M.~V.}\ \bibnamefont {Chekhova}}, \ and\
  \bibinfo {author} {\bibfnamefont {P.~{\relax St}.~J.}\ \bibnamefont
  {Russell}},\ }\href {\doibase 10.1103/PhysRevLett.115.143602} {\bibfield
  {journal} {\bibinfo  {journal} {Physical Review Letters}\ }\textbf {\bibinfo
  {volume} {115}},\ \bibinfo {pages} {143602} (\bibinfo {year}
  {2015})}\BibitemShut {NoStop}%
\bibitem [{\citenamefont {Loudon}\ and\ \citenamefont
  {Knight}(1987)}]{LOUDON1}%
  \BibitemOpen
  \bibfield  {author} {\bibinfo {author} {\bibfnamefont {R.}~\bibnamefont
  {Loudon}}\ and\ \bibinfo {author} {\bibfnamefont {P.~L.}\ \bibnamefont
  {Knight}},\ }\href@noop {} {\bibfield  {journal} {\bibinfo  {journal}
  {Journal of Modern Optics}\ }\textbf {\bibinfo {volume} {34}},\ \bibinfo
  {pages} {709} (\bibinfo {year} {1987})}\BibitemShut {NoStop}%
\bibitem [{Note1()}]{Note1}%
  \BibitemOpen
  \bibinfo {note} {By definition, the photon creation and annihilation
  operators in $g^{(2)}$ should follow normal ordering. When the number of
  photons is high, it is possible to drop normal ordering.}\BibitemShut {Stop}%
\bibitem [{\citenamefont {Abram}\ \emph {et~al.}(1986)\citenamefont {Abram},
  \citenamefont {Raj}, \citenamefont {Oudar},\ and\ \citenamefont
  {Dolique}}]{PhysRevLett.57.2516}%
  \BibitemOpen
  \bibfield  {author} {\bibinfo {author} {\bibfnamefont {I.}~\bibnamefont
  {Abram}}, \bibinfo {author} {\bibfnamefont {R.~K.}\ \bibnamefont {Raj}},
  \bibinfo {author} {\bibfnamefont {J.~L.}\ \bibnamefont {Oudar}}, \ and\
  \bibinfo {author} {\bibfnamefont {G.}~\bibnamefont {Dolique}},\ }\href
  {\doibase 10.1103/PhysRevLett.57.2516} {\bibfield  {journal} {\bibinfo
  {journal} {Physical Review Letters}\ }\textbf {\bibinfo {volume} {57}},\
  \bibinfo {pages} {2516} (\bibinfo {year} {1986})}\BibitemShut {NoStop}%
\bibitem [{gla()}]{glass1}%
  \BibitemOpen
  \href@noop {} {\enquote {\bibinfo {title} {Schott optical glass data sheets,
  2012},}\ }\bibinfo {howpublished}
  {\url{http://refractiveindex.info/download/data/2012/schott_optical_glass_collection_datasheets_dec_2012_us.pdf
  }}\BibitemShut {NoStop}%
\bibitem [{Note2()}]{Note2}%
  \BibitemOpen
  \bibinfo {note} {See Supplemental Material at the publisher's website for
  details on the calibration of the spectrometer and on the effect of phase
  fluctuations on the shape of the frequency spectrum.}\BibitemShut {Stop}%
\bibitem [{\citenamefont {Iskhakov}\ \emph {et~al.}(2016)\citenamefont
  {Iskhakov}, \citenamefont {Lemieux}, \citenamefont {P{\'e}rez}, \citenamefont
  {Boyd}, \citenamefont {Chekhova},\ and\ \citenamefont {Leuchs}}]{ISKHAKOV3}%
  \BibitemOpen
  \bibfield  {author} {\bibinfo {author} {\bibfnamefont {T.~{\relax Sh}.}\
  \bibnamefont {Iskhakov}}, \bibinfo {author} {\bibfnamefont {S.}~\bibnamefont
  {Lemieux}}, \bibinfo {author} {\bibfnamefont {A.~M.}\ \bibnamefont
  {P{\'e}rez}}, \bibinfo {author} {\bibfnamefont {R.~W.}\ \bibnamefont {Boyd}},
  \bibinfo {author} {\bibfnamefont {M.~V.}\ \bibnamefont {Chekhova}}, \ and\
  \bibinfo {author} {\bibfnamefont {G.}~\bibnamefont {Leuchs}},\ }\href@noop {}
  {\bibfield  {journal} {\bibinfo  {journal} {Journal of Modern Optics}\
  }\textbf {\bibinfo {volume} {63}},\ \bibinfo {pages} {64} (\bibinfo {year}
  {2016})}\BibitemShut {NoStop}%
\bibitem [{\citenamefont {Zou}\ \emph {et~al.}(1991)\citenamefont {Zou},
  \citenamefont {Wang},\ and\ \citenamefont {Mandel}}]{zou1991induced}%
  \BibitemOpen
  \bibfield  {author} {\bibinfo {author} {\bibfnamefont {X.~Y.}\ \bibnamefont
  {Zou}}, \bibinfo {author} {\bibfnamefont {L.~J.}\ \bibnamefont {Wang}}, \
  and\ \bibinfo {author} {\bibfnamefont {L.}~\bibnamefont {Mandel}},\
  }\href@noop {} {\bibfield  {journal} {\bibinfo  {journal} {Physical Review
  Letters}\ }\textbf {\bibinfo {volume} {67}},\ \bibinfo {pages} {318}
  (\bibinfo {year} {1991})}\BibitemShut {NoStop}%
\bibitem [{Note3()}]{Note3}%
  \BibitemOpen
  \bibinfo {note} {Note that by selecting a single spatial mode, one does not
  yet select a single frequency mode, and vice versa.}\BibitemShut {Stop}%
\end{thebibliography}%

\end{document}